\documentclass{article}
\usepackage{sis_conf}
\usepackage{graphicx}

\begin{document}

\long\def\comment#1{}
\newcommand{\etc}{{\em etc}}
\newcommand{\eg}{{\em e.g.}}
\newcommand{\ie}{{\em i.e.}}
\newcommand{\ea}{{\em et al.}}
\newcommand{\noi}{\noindent}
\newcommand{\remove}[1]{}  

\newenvironment{itm}{\begin{itemize}\setlength{\itemsep}{-0.02in plus 0in}}{\end{itemize}}

\title{Modeling and Mathematical Analysis of Swarms of Microscopic Robots}
\name{Aram Galstyan$^1$, Tad Hogg$^2$, Kristina Lerman$^1$}
\address{1. USC Information Sciences Institute\\
4676 Admiralty Way\\
Marina del Rey, CA 90292, USA \\
 2. HP Labs \\
  1501 Page Mill Road, MS 1U-19 \\
   Palo Alto, CA 94304, USA
   }

\maketitle




\begin{abstract}
The biologically-inspired swarm paradigm is being used to design
self-organizing systems of locally interacting artificial agents.
A major difficulty in designing swarms with desired
characteristics is understanding the causal relation between
individual agent and collective behaviors. Mathematical analysis
of swarm dynamics can address this difficulty to gain insight into
system design. This paper proposes a  framework for mathematical
modeling of swarms of microscopic robots that may one day be
useful in medical applications. While such devices do not yet
exist, the modeling approach can be helpful in identifying various
design trade-offs for the robots and be a useful guide for their
eventual fabrication. Specifically, we examine microscopic robots
that reside in a fluid, for example, a bloodstream, and are able
to detect and respond to different chemicals. We present the
general mathematical model of a scenario in which robots locate a
chemical source. We solve the scenario in one-dimension and show
how results can be used to evaluate certain design decisions.

\end{abstract}

\section{Introduction}

Swarm Intelligence is a relatively new concept for designing
multi-agent systems~\cite{bonabeau99}. Swarms are decentralized
systems composed of many simple agents with no central controller.
Instead, swarms are {\em self-organizing}: constructive
collective\remove{ ({\em macro\-scopic})} behavior emerges from
local\remove{ ({\em micro\-scopic})} interactions among the agents
and between agents and their environment. Self-organization is
ubiquitous in nature: Aggregation into multi-cell organisms in
{\em Dictyostelium discoideum} colony, trail formation in ants,
hive building by bees, and mound construction by termites, are a
few examples. The success of these organisms has inspired the
design of distributed problem-solving systems.

A major difficulty in designing self-organizing swarms with
desired characteristics is understanding the causal relation
between individual agent characteristics and collective behavior.
One approach is experiments with real agents, {\em e.g.}, robots,
to evaluate behavior under real conditions. However, fabricating
enough agents to exhibit swarm behavior is often too expensive or
not feasible with current technology. Simulations, such as
sensor-based simulations of robots, are usually much faster and
less costly than experiments. But these require good models of the
robots and their environments, which are often not yet available.
Moreover, simulations still require many repetitions to
systematically examine consequences of various choices for
individual agent behaviors.

Mathematical modeling and analysis offer an alternative to the
time-consuming and costly experiments and simulations.
Mathematical analysis  can be used to study swarm dynamics,
predict long-term behavior of even very large systems, and gain
insight into system design: {\em e.g.}, how individual agent
characteristics affect the swarm behavior. Such analysis can
suggest parameters that optimize group performance, prevent
instabilities, \etc. These particular choices could then be
examined in more detail through simulations or experiments.

This mathematical approach is particularly useful for preliminary
evaluation of design trade-offs for robots that cannot yet be
built and will operate in poorly characterized environments. As
one such application, robots with sizes comparable to bacteria
(``nanorobots'') could provide many novel capabilities through
their ability to sense and act in microscopic environments. Of
particular interest are medical applications, where robots and
nanoscale-structured materials inside the body could significantly
improve disease diagnosis and
treatment~\cite{freitas99,morris01,nih03,keszler01}. Typically,
each such robot will have dimensions of a few microns so could
readily travel through even the tiniest blood vessels, will be
able to communicate with other robots (e.g., using ultrasound)
over distances of about $100$ microns, and will consume in the
range of $1-1000$ picowatts.

An example task for such robots operating in the bloodstream is to
aggregate and form structures at locations with specific patterns
of chemicals. Initially, as robots move passively through the
fluid, monitoring for the chemical pattern, they need to avoid
disturbing the flow. Thus they will be at low density, i.e., with
typical separation between neighboring robots many times larger
than their size. On the other hand, to affect the environment,
many robots will need to act together (e.g., to block flow in
blood vessels feeding populations of cancer cells or to aid in
repairing damaged tissues by forming suitable scaffolding). Thus
the swarm must change to a higher density of robots, in spite of
limited communication ranges. In this denser configuration,
communication among neighbors is easier, and simple local controls
allow forming a variety of structures either static or dynamically
responding to changing forces in the environment, as proposed in
the context of large self-reconfigurable
robots~\cite{bojinov02,salemi01}. To quantify this example, a
swarm could consist of around $10^{12}$ machines, vastly more than
in swarms of larger devices. Even such a large number of machines
would have a total mass of only about one gram, and a occupy less
than $0.1\%$ of the typical blood volume of an adult, compared to
the $20-40\%$ occupied by cells.

While fabricating such machines is beyond the ability of current
technology, the rapid progress in engineering na\-no\-scale
devices should eventually enable their production. For example,
ongoing
development~\cite{\remove{collier99,craighead00,}howard97,fritz00,montemagno99\remove{,soong00}}
of molecular-scale electronics, sensors and motors provides
components for such robots, though these cannot yet be assembled
into large numbers of complete systems. Programmable
bacteria~\cite{\remove{weiss99,}weiss00} give large numbers of
functioning microscopic systems, and separate DNA computers can
respond to Boolean combinations of chemicals in their
environment~\cite{benenson04}. However relying on biological
organisms restricts the robots' material properties and, by using
protein synthesis to perform logic operations, limits the
computational complexity and speed of control programs.

Beyond the challenge of fabricating the robots is designing
 and analyzing controls suitable for their physical environments, tasks and
capabilities, whose features differ considerably from those of
larger robots. Fortunately, we can evaluate control methods prior
to building the robots with theoretical
studies~\cite{drexler92,freitas99,requicha03} characterizing robot
capabilities and their task environments. Moreover, suitable
control programs can compensate for some hardware limitations
(e.g., limited available power or sensor accuracy). Thus studies
of control approaches can identify useful tasks for robots with
limited hardware capabilities, as may arise before the fabrication
technology is fully mature.

%
This paper extends the stochastic processes framework developed
for modeling teams of larger robots and that has been applied
successfully to a number of robotic
applications~\cite{lerman04sab}. Our work complements other
theoretical studies of microscopic
robots~\cite{cavalcanti03,requicha03,gazi04} by providing a
methodology for studying collective swarm behavior. It can also be
applied to larger robots that implement chemotaxis as the basis
for movement in an external field~\cite{dhariwal04}. While
comparison with microscopic experiments must await the technical
feasibility of fabricating microscopic robots, our mathematical
model can be used to evaluate various design trade-offs for the
robot capabilities that can achieve the same swarm performance and
can be a useful guide to eventual fabrication by allowing
designers to focus on combinations of hardware capabilities
easiest to fabricate.

Below we present a general model of the dynamics of a swarm of
microscopic robots that are interacting with an external field. In
Section~\ref{sec:scenario} we motivate the approach by describing
a medically relevant scenario that considers a swarm of
microscopic robots moving in a fluid to localize a chemical
source. In Section~\ref{sec:results} we solve a one-dimensional
model and analyze different design choices. We conclude the paper
with a discussion of the approach and results.

\section{Stochastic Theory of Spatially-dependent Processes}
\label{sec:model}
Previously, we developed a framework based on stochastic Master
Equation for mathematical analysis of homogeneous robotic
swarms~\cite{Huberman88Eco,lerman02nasa}. Specifically, consider
an agent whose controller is an automaton consisting of $L$
\textit{local} states. We then assume the overall system is
sufficiently well-described by an occupancy vector ${\bf n} =
(n_1, n_2,...n_L)$, where $n_k$ is the fraction of agents in state
$k$. That is, we need only consider the number of agents in each
state rather than tracking their individual identities. In
designing systems where typical or average behavior is of primary
interest, it is useful to characterize the system by a probability
distribution $P({\bf n},t)$, which is the probability the system
is in configuration ${\bf n}$ at time $t$, and evolves according
to the Master Equation~\cite{VanKampen}. This distribution
characterizes the behaviors of an ensemble or multiple independent
instances of the swarm. When most such instances have nearly the
same behavior,\footnote{Consider, for example, a swarm consisting
of a large number of agents, each independently executing the same
controller. Then, by the law of large numbers, behaviors of
different instances of this swarm would be nearly identical. }
fluctuations among different instances can be neglected, and the
average occupancies $\bar{n}_k$ provide a sufficient
characterization of the system. The average fraction of agents in
state $k$ then evolves according to
\begin{equation}
\label{eqn-rate-1} {d \bar{n}_k \over {d t}} = \sum_{j} w_{jk}
\bar{n}_{j} -  \bar{n}_k  \sum_{j}w_{kj}
\end{equation}
with $w_{jk}$ the transition rates between the states $j$ and $k$.
Note than in general $w_{jk}$ might depend on the fraction of
agents in particular states, i.e., average occupancy vector
$\bar{{\bf n}}$. Eq.~\ref{eqn-rate-1} governs the evolution of the
collective behavior of the system.
\remove{In particular, the steady state solution $\bar{n}_k(t
\rightarrow \infty)$ describes the system's long term behavior.}

While this approach works well for many spatially uniform
systems~\cite{lerman04sab}, it is too coarse-grained for systems
with a spatial correlation in agents' interactions. Thus, it is
not sufficient to describe, for example, an ant-like swarm where
agents interact through evolving chemical fields or microscopic
robots monitoring chemicals released into a fluid. These
situations require a generalization of the Master Equation, in
which each robot not only has a discrete controller state $k$ but
also a continuous coordinate $\bf{x}$ (i.e., its spatial
location). As with the original formulation, we suppose the number
of agents in each state is sufficient to determine the collective
behavior of interest. Because ${\bf x}$ is a continuous variable,
these counts become densities leading us to introduce
$\bar{n}_k({\bf x},t)$ as the average robot fraction density in
state $k$ at location ${\bf x}$ and time $t$. Thus a small volume
$\Delta x$ around location ${\bf x}$ contains, on average, the
fraction $\bar{n}_k({\bf x},t)\Delta x$ of the robots in the
system.

Let us consider a system where agents interact with the
environment through a certain external chemical field. Let us also
assume that agents are able to interact through stigmergy by
releasing a special chemical into the environment that we call
\textit{communicative signal}. We denote  $\rho({\bf x}, t)$ and
$c({\bf x},t)$  concentration of the chemical and communicative
signal, respectively, at point ${\bf x}$ at time $t$.  Then we can
write down the generalized rate equation as follows:
\begin{eqnarray}
\label{eqn-rate-2} {\partial \bar{n}_k({\bf x},t) \over {\partial
t}} &=& \int
d{\bf x}' \sum_{j} w_{jk}({\bf x},{\bf x}';\rho,c) \bar{n}_{j}({\bf x}',t)  \\
 &-&   \bar{n}_k({\bf x},t) \int d{\bf x}'
\sum_{j}w_{kj}({\bf x},{\bf x}';\rho,c)\,. \nonumber
\end{eqnarray}
Now the transition rates $w_{jk}$ depend not only the state
indices $j$ and $k$ and occupation vector  but also on the spatial
coordinates and concentration of the chemical at the corresponding
points. Note also that we have included the dependence of the
transition rates on ${\bf x}$ and ${\bf x}'$ explicitly to account
for agents' kinematics even in the absence of chemical and
communicative concentrations (e.g., to describe  freely diffusing
agent).

The transition rates $w_{jk}({\bf x},{\bf x}';\rho({\bf x}),c({\bf
x}))$ summarize the behaviors of the individual robots. For
example, the robot's internal state could change when it detects a
chemical concentration above a predetermined threshold. In
applications of microscopic robots, such detections will often
involve a relatively small number of molecules (e.g., on the order
of tens) so stochastic fluctuations in the number detected will be
a significant source of sensor noise and can be included as a
range of new states with nonzero transition rates. Communication
among nearby robots, e.g., with ultrasound, allows the robots to
reduce noise in estimating chemical gradients and hence perform
better than individual robots or bacteria~\cite{adler66}, but at a
cost of additional power use for the communication. Robot motion,
either moving passively with the fluid or using powered
locomotion, e.g., to follow chemical concentration gradients, also
contributes to the transitions.

While Equation~\ref{eqn-rate-2} is a general description of the
overall system behavior, it is too complex in its present form to
be useful. Fortunately, it can be simplified considerably into a
more intuitive form  by noting that in many physically realistic
situations agents' motion can be decoupled from  state
transitions, so that the transition rate can be represented as
\begin{eqnarray}
\label{eq:w} w_{jk} &=& \delta_{jk} W_k({\bf x},{\bf x}';\rho({\bf
x}),\rho({\bf x}'),c({\bf x}),c({\bf x}')) \nonumber \\&+&
\delta(x-x')w_{jk}(\rho({\bf x}),c({\bf x}))\,,
\end{eqnarray}
where $\delta_{jk}$ is Kroenecker's symbol\footnote{Kroenecker's
symbol is defined as follows: $\delta_{ij}=1$ if $i=j$ and
$\delta_{ij}=0$, $i \neq j$.} and $\delta(x)$  is its continuous
analogue $\delta$--function.  In other words, during a transition
between two discrete states we neglect  the change in robot's
position. In Eq.~\ref{eq:w} $W_k$ is an appropriately chosen
kernel that describes agents' motion (as index $k$ indicate, it
can be different for each state), while the second term describes
transition between discrete states.

Equation~\ref{eq:w} allows us to separate transition function into
terms with purely spatial transitions and terms with purely state
transitions. Indeed, using Eq.~\ref{eq:w} we can decouple the
agents' kinematics from the state transitions between discrete
state and rewrite Eq.~\ref{eqn-rate-2} as follows:
\begin{eqnarray}
\label{eqn-rate-3} {\partial \bar{n}_k({\bf x},t) \over {\partial
t}} &=& \mathcal{L}_k \bar{n}_k ({\bf x},t) + \sum_{j} w_{jk}(\rho,c) \bar{n}_{j}({\bf x},t)~~~ \\
 &-&   \bar{n}_k({\bf x},t)
\sum_{j}w_{kj}(\rho,c) \nonumber
\end{eqnarray}
Here $\mathcal{L}_k$ is an operator (specified below) that
describes the motion of agents in state $k$. The second and third
terms in Eq.~\ref{eqn-rate-3} describe agents state transitions.
Note that now $w_{jk}(\rho,c)$ depends on spatial coordinates
indirectly, through concentration $\rho({\bf x},t)$ and $c({\bf
x},t)$. When the concentrations $\rho$ and $c$ are constants,
Eq.~\ref{eqn-rate-1} is recovered by integrating
Eq.~\ref{eqn-rate-3} over the spatial coordinates ${\bf x}$,
assuming that $\mathcal{L}_k$ preserves the number of agents in
state $k$ (e.g., no absorbing boundaries) so that integral over
the first term in Eq.~\ref{eqn-rate-3} vanishes.

To specify the operators $\mathcal{L}_k$, we note that for the
particular environment we are interested in, (i.e., microscopic
robots operating in a  fluid) robots' motion can be described by a
diffusion equation~\cite{VanKampen}. In this paper, we study
chemotactic robots that respond to a chemical and signalling
fields by propelling themselves in the direction of increasing
concentration. This capability is modeled after bacterial
chemotaxis which allows these single cell organisms to efficiently
move towards food sources and away from noxious sources. Although
in some cases the exact derivation from the microscopic transition
rates is feasible, if very involved (see, for example,
\cite{Othmer02,Erban04} for treatment of bacterial chemotaxis
which can be treated as a biased random walk), chemotaxis in a
chemical concentration field $\rho( {\bf x},t)$ is usually
introduced into the rate equations phenomenologically by
postulating a chemotactic velocity as $V_{D} = \eta_{\rho} \nabla
\rho( {\bf x},t)$, where $\eta_{\rho}$ is the so called
chemotactic sensitivity (which may itself depend on $\rho$).  One
can then write for operators $\mathcal{L}_k$
\begin{equation}
\label{eqn-op-rob} \mathcal{L}_k =  D_{k} \nabla^2 - {\bf v} \cdot
\nabla - \nabla \cdot [{\bf V}_{D}^{\rho}(\rho, \nabla \rho) +
{\bf V}_{D}^{c}(c, \nabla c)]
\end{equation}
Here, $D_{k}$ is the diffusion coefficient of agents in state $k$
assumed to be a constant, ${\bf v}$ is the flow velocity, and
${\bf V}_{D}^{\rho}$ and ${\bf V}_{D}^{c}$ are the chemotaxis
drift velocities of robots due to concentration gradients of the
 chemical and the communicative signal, respectively.

To proceed further, we should also define how the chemical and
concentration fields evolves in time. As an example relevant for
microscopic robots, we consider the evolution of this fields in a
moving fluid in which the robots operate. The evolutions of
$\rho({\bf x},t)$ $c({\bf x},t)$ are governed by the diffusion
equation:
\begin{eqnarray}
\label{eqn-chemical} {\partial \rho \over {\partial t}} &=&
D_{\rho} \nabla^2\rho - {\bf v} \cdot \nabla \rho -
\gamma_{\rho} \rho+Q_{\rho}({\bf x},t) \\
\label{eqn-signal} {\partial c \over {\partial t}} &=& D_{c}
\nabla^2 c - {\bf v} \cdot \nabla c - \gamma_{c}c+\sum_k
q_{k}\bar{n}_k({\bf x},t)
\end{eqnarray}

In Eq.~\ref{eqn-chemical} the terms on the right describe,
respectively, the diffusion of the chemical (with a diffusion
constant $D_{\rho}$), the advection of the chemical due to fluid
motion with velocity ${\bf v}$, the decay of the chemical at rate
$\gamma_{\rho}$, and its deposition by sources with intensity
profile $Q_{\rho}({\bf x},t)$. Terms in Eq.~\ref{eqn-signal} have
similar meaning, except the deposition rates of signalling
chemical is proportional to the fraction of agents in state $k$,
$\bar{n}_k({\bf x},t)$ (note that, generally speaking,  the
coefficients $q_k$ themselves  depend on the fraction of agents in
state $k$).  The parameters in this equation could, in general,
depend on space and time, as well as the location of the robots
(e.g., a sufficiently high concentration of the robots could
significantly affect the fluid flow). For simplicity, we will
treat them as constants. For microscopic robots, fluid motions
will usually be at very low Reynolds number so the fluid flow will
be laminar with the velocity ${\bf v}$ changing smoothly with
location. Viscous forces dominate the motions of such robots with
requirements for locomotion mechanisms and power use quite
different from experiences with larger robots~\cite{purcell77}.

Note that the system of Eqs.~\ref{eqn-rate-3}, \ref{eqn-chemical}
and~\ref{eqn-signal} can be easily generalized to handle multiple
chemicals with different diffusion coefficients. Another
generalization would allow a variable number of agents in the
system, in which case we would add a variable for the total number
of agents in the system at time $t$ to the variables introduced
above describing the fraction in various states.

Equations~\ref{eqn-rate-3}--\ref{eqn-signal}  together with
appropriately chosen bo\-un\-dary conditions describe time
evolution of the system. In the next section we use this
formulation for a prototypical task: finding and localizing at the
source of a chemical  released into a flowing fluid.

\section{Target Localization with Microscopic Robots}
\label{sec:scenario}

Let us consider a D--dimensional volume with multiple targets that
release certain chemical into the environment. The task of the
microscopic swarm is to aggregate at these targets in order to
carry out some actions in the vicinity of the targets. This
capability is fundamental to many medical applications envisioned
for these microscopic robots. For example, the volume of fluid may
be a blood vessel that has been damaged. Robots are required to
aggregate at the injury site in order to assist in healing,
forming clots, etc.

We consider a simple robot controller that on a high level can be
thought to consist of 3 discrete states described below:

\begin{description}

\item[State 1 (search):] Do a biased random walk  in the
 direction of the communicative signal concentration gradient.

\item[State 2 (communicate):] Move  towards the chemical so\-urce
following the concentration gradient of the target chemical and
release communicative signal to other robots.

\item[State 3 (disperse):] Move away from the target in the
direction opposite to the target chemical's concentration gradient
for some specified time $\tau$.

\end{description}

To fully specify a robot's behavior, we also have to describe the
transitions between these states. The robots start out in State 1,
the searching for targets using random diffusive motion and
following the gradient of the communicative signal. Once the
concentration of the target chemical at a certain point in space
is sufficiently high the robot at that point will switch to the
State 2: it will start moving towards regions of high
concentration (using biased diffusion or gradient following) while
releasing a new chemical which acts as a communication signal to
attract other robots. With some probability (that can be fixed, or
dependent on the concentration of the robots at the source),
robots in the State 2 will switch to State 3, where they will
disperse from the source, moving
 in the direction opposite to the gradient.
Finally, robots in the State 3 will switch to the searching state
with probability $1/\tau$. The last behavior ensures that robots
will not be stuck at local maxima of the chemical potential.

Let denote by $n_1(\textbf{x})$, $n_2(\textbf{x})$,
$n_3(\textbf{x})$ the fraction of robots in each state at point
$\textbf{x}$, with normalization condition
$$\int d \textbf{x} \big(
n_1(\textbf{x}) + n_2(\textbf{x})+n_3(\textbf{x})\big)=1.$$ Let
$\rho(\textbf{x})$ and $c(\textbf{x})$ be the concentrations of
the chemical released from the targets and the communicative
signal released by robots in State 2. We also denote by ${\bf
V}_{D}^{\rho}$ and ${\bf V}_{D}^{c}$ the robots' drift velocity in
the concentration gradients of chemical (released by the targets)
and communicative signal (released by the robots), respectively.
Then the set of equations describing the evolution of the system
is as follows:
\begin{eqnarray}
\label{eq:n1} \frac{\partial n_1}{\partial t}  & = &  D_1\nabla^2
n_1 - \textbf{v}\cdot \nabla n_1 - \nabla \cdot [{\bf V}_{D}^{c}
n_1]  \nonumber\\
&-&n_1F(\rho) + \frac{n_3}{\tau} \\
\label{eq:n2} \frac{\partial n_2}{\partial t}  & = &  D_2 \nabla^2
n_2- \textbf{v}\cdot \nabla n_2 - \nabla \cdot [{\bf
V}_{D}^{\rho} n_2]  \nonumber \\
&+& n_1F(\rho) - G(n_2,\rho,c)n_2  \\
\nonumber\\
\label{eq:n3} \frac{\partial n_3}{\partial t} & = & D_3\nabla^2
n_3 - \textbf{v}\cdot \nabla n_3 + \nabla \cdot [{\bf
V}_{D}^{\rho} n_3]  \nonumber \\
&+&G(n_2,\rho,c)n_2 -\frac{n_3}{\tau}
\end{eqnarray}
where $F(\rho)$ is the concentration--dependent transition rate
from State $1$ to State $2$, $G(n_2;\rho;c)$ is the transition
rate from State $2$ to State $3$, and $1/\tau$ is the probability
that a robot in State $3$ will switch to State $1$.

Equations \ref{eq:n1}--\ref{eq:n3} have a simple intuitive
interpretation. The first two terms in Eq.~\ref{eq:n1} describe
robots motion in State 1: diffusive searching and following
communicative signal, if present. The third term describes the
drift in the flow. The fourth term describes transitions to State
2 at the rate $F(\rho)$, which depends on the concentration of the
target field. The last term describes transition of robots from
State 3 to State 1 after the robots have moved in the direction
opposite to the concentration gradient for a period of time
$\tau$. $G(n_2, \rho, c)$ is the rate at which robots transition
from State 2 to State 3, and it could in principle depend on the
local concentrations of the gradients, as well as the number of
agents present at the target site, for example, when presence of a
certain minimum number of robots is required for executing an
action.

We have to complement these three equations with two more to
account for the evolution of chemicals $\rho$ and $c$, that are
obtained from
Equations~\ref{eqn-chemical-signal},\ref{eqn-op-chemical}
and~\ref{eqn-op-signal} as follows:
\begin{eqnarray}
\label{eq:rho} \frac{\partial \rho}{\partial t} & = &
D_{\rho}\nabla^2 \rho -\textbf{v} \cdot \nabla \rho +
\sum_{i=1}^{M}Q_i\delta(\textbf{x} - \textbf{x}_i)- \gamma_{\rho}
\rho~~~~~~\\
\label{eq:c} \frac{\partial c}{\partial t} & = & D_{c}\nabla^2 c
-\textbf{v} \cdot \nabla c  + q_cn_2 - \gamma_c c
\end{eqnarray}
In Eq.~\ref{eq:rho} ${\bf x}_i$-s, $i=1,2,..M$ are the locations
of the target sources, $Q_i$ is the intensity of source $i$, and
$\gamma_{\rho}$ is the decay rate of the target chemical.
Similarly, in Eq.~\ref{eq:c} $q_c$ is the intensity of
communication signal released by a robot in State 2, while
$\gamma_c$ is the decay rate of the signal.

\section{Results for a Simplified 1D Scenario}
\label{sec:results}

In this section we present results for a  $1D$ geometry and a
single target scenario. We consider the case when the liquid flow
is very slow compared to other time scales so we can set ${\bf v}
=0$. Also, since there is only one target, we neglect the third
(dispersing) behavior so that two possible states are State 1
(``search'') and State 2 (``communicate''). The target is located
at $x=1$ and serves as a point source for the chemical. We assume
that the diffusion of the chemical happens much faster compared to
robots' diffusion, and it quickly reaches its steady state
profile. Hence, the equation for evolution of $\rho(x,t)$ can be
solved separately, with a solution
\begin{equation}
\rho(x,\infty) \equiv \rho(x) = Q_0
e^{-\sqrt{\gamma_{\rho}/D_{\rho}}(1-x)}, 0\le x\le 1.
\end{equation}
For the results presented here we used $Q_0=0.1$, $D_{\rho}=0.2$
and $\gamma_{\rho}=0.5$.

 All robots start at State $1$ and are
initially localized at $x=0$. We assume that a transition from
State $1$ to State $2$ happens whenever a robot in State $1$
detects the target's chemical above a certain threshold level
$\rho_0$, so that the transition rate is $F(\rho) = \theta(\rho -
\rho_0 )$, where $\theta(x)$ is the step function, $\theta(x)=1$
if $x\ge 0$ and $\theta(x)=0$, $x < 0$. While in State 2, robots
move in the chemical gradient with a constant drift velocity $V_D$
and release a communicative signal with intensity $q_c$.

To proceed further, we need to specify the dependence of the drift
velocity in State 1 on the concentration of communicative signal
$c$. Again, we assume that once a robot detects  communicative
signal above certain threshold $c_0$, it propels itself through
the fluid in the direction of the gradient with a constant drift
velocity $V_D$. Then the dynamics of the system is described by
the following system of equations:
\begin{eqnarray}
\label{eqb:n1} \frac{\partial n_1}{\partial t}   &=&
D_n\frac{\partial^2 n_1}{\partial x^2} - V_D\theta(c-c_0)\frac{\partial n_1}{\partial x} - F(\rho)n_1~~~\\
\label{eqb:n2} \frac{\partial n_2}{\partial t}   &=&
D_n\frac{\partial^2 n_2}{\partial x^2}  - V_D\frac{\partial n_2}{\partial x}+ F(\rho) n_1~~~~\\
\label{eqb:c} \frac{\partial c}{\partial t}  & =&
D_c\frac{\partial^2 c}{\partial x^2} +  q_cn_2 - \gamma_c c ~~~
\end{eqnarray}

To study the effect of different design parameters on aggregation
behavior of the robots at the target, we solved the system
Eq.~\ref{eqb:n1}--~\ref{eqb:c} numerically. We used the following
parameters (in dimensionless units): $D_n=0.01$, $D_c=0.05$,
$V_D=0.1$, $q_c = 0.1$, $\gamma_c = 0.01$. For the detection
thresholds we used $c_0=0.001$ and $\rho_0 = 0.01$, the later
assuring that that robots detect the chemical approximately midway
in the interval $[0,1]$. We used reflective  boundary conditions
for $n_1$ and $n_2$,  $\partial n_1/\partial x|_{0,1} = \partial
n_2/\partial x|_{0,1}=0$, and absorbing boundary conditions for
$c$, $c(0)=c(1)=0$.

\begin{figure}[tbhp]
\begin{tabular}{c}
\includegraphics[width=0.4\textwidth]{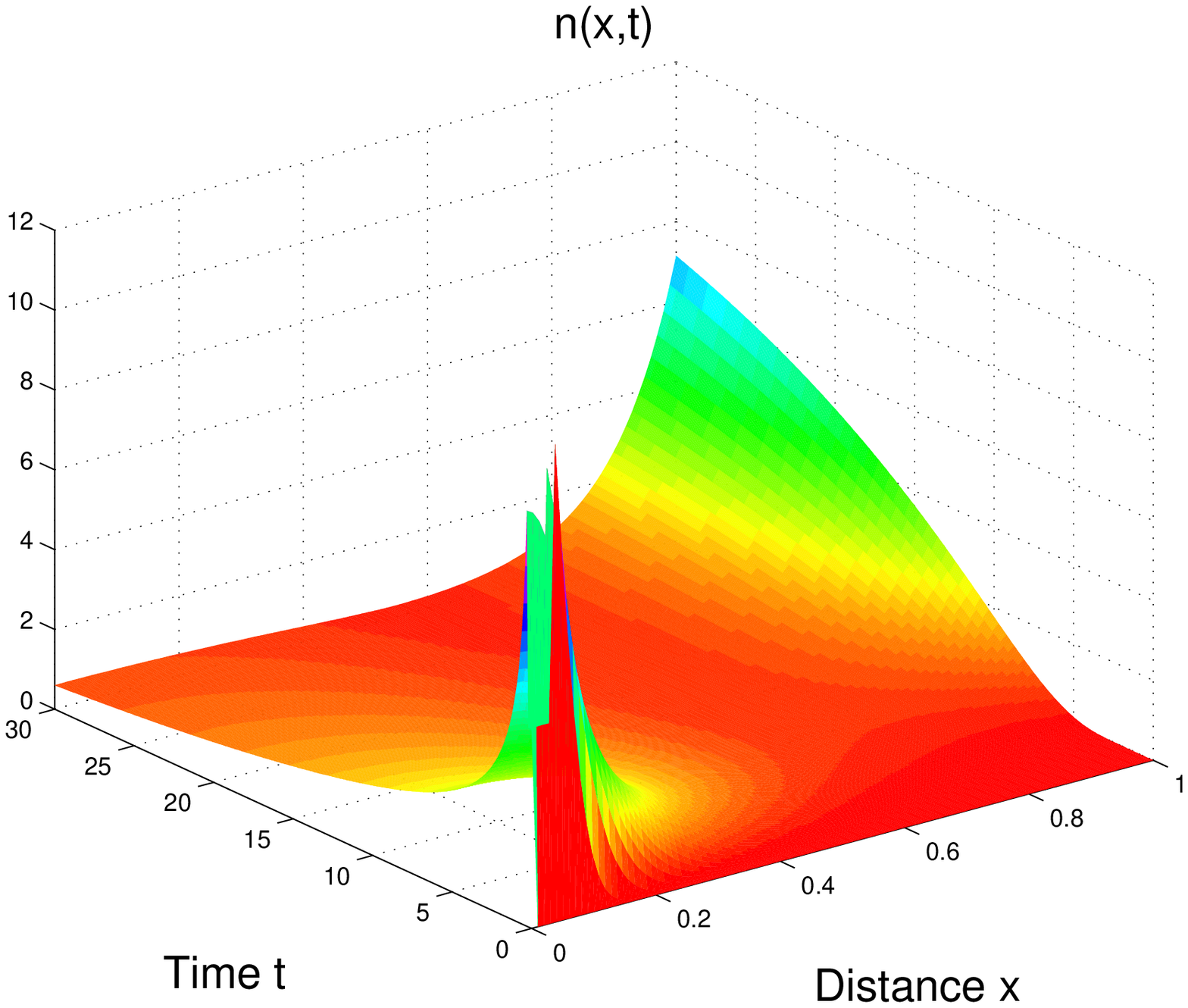} (a)\\
\includegraphics[width=0.4\textwidth]{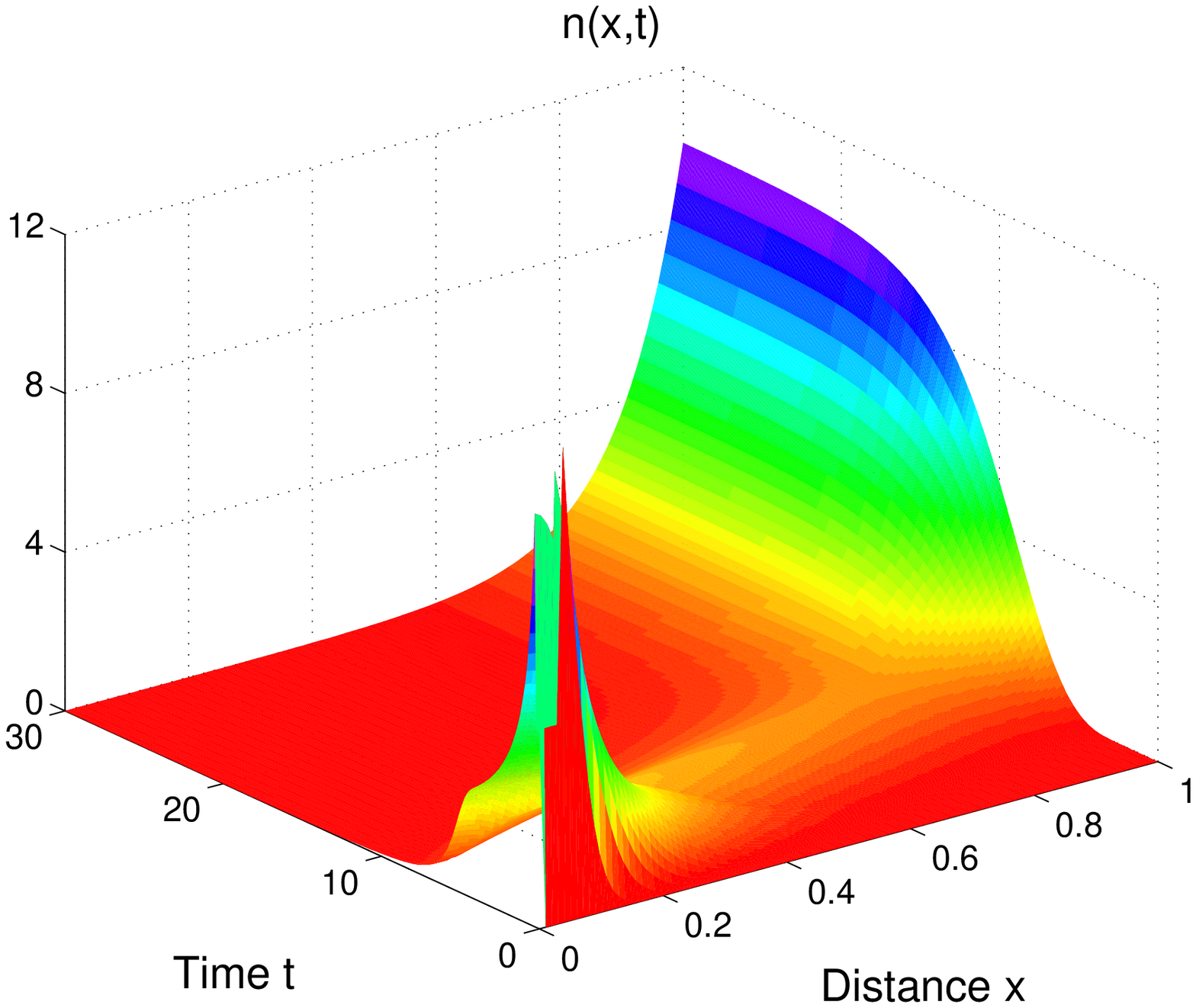} (b)
\end{tabular}
\caption{Time evolution of robot densities without communication
(a) and with communication (b)}\label{fig:3d}
\end{figure}

\begin{figure}[tbhp]
\includegraphics[width=0.4\textwidth]{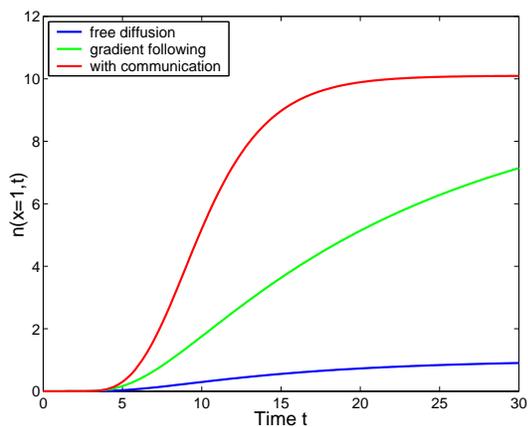}
\caption{Time evolution of robot densities at $x=1$ for three
different strategies}\label{fig:at1}
\end{figure}

In Fig.~\ref{fig:3d} we plot the spatia--temporal evolution of
robots' densities with and without communication. Clearly, the
density peak at $x=1$ is stronger for the system with
communicative behavior. This suggests that communication indeed
helps the robots to aggregate better. In addition, the aggregation
process with communication happens faster than without
communication. This is also shown in Fig.~\ref{fig:at1}, where we
plot the density of robots at $x=1$ as a function of time for
three different cases: free diffusion\footnote{Note that the
absence of aggregation for free diffusing robots is due to
reflective boundary conditions at the source for $n_1$ and $n_2$.
If one employs absorbing boundary conditions instead, robots will
demonstrate aggregative behavior even with free diffusion.}
($V_D=0$), gradient following without communication ($V_D\neq 0,
q_c=0$), and gradient following with communication ($V_D,q_c\neq
0$). As it can be seen from Fig.~\ref{fig:at1}, the systems with
gradient following and communicative behavior do  demonstrate
aggregative behavior, and it is more pronounced for the system
with communication. For instance, at time $t=10$ the robot density
at $x=1$ and with communication is  more than $3$ times higher
than in the non--communicating case.

\begin{figure}[tbhp]
\includegraphics[width=0.4\textwidth]{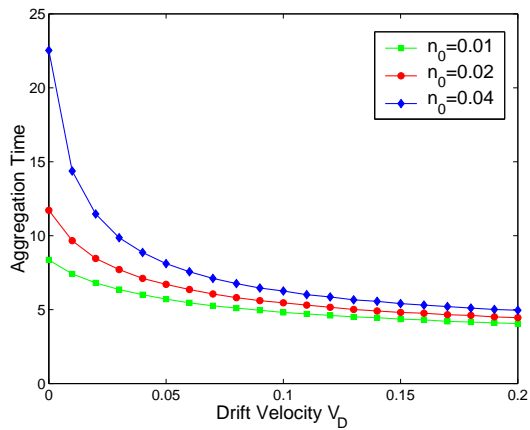}
\caption{Aggregation time as a function of drift velocity, for
three different values of $n_0$.}\label{fig:conv}
\end{figure}

One of the design objectives is to have robots aggregate at the
target fast enough, while at the same time not dissipating too
much power due to the propelling. To examine this tradeoff, let us
consider the dependence of the aggregation time (defined as time
needed for fraction $n_0$ of robots to reach the vicinity of the
target which we define as the interval $[0.95,1]$,) on the drift
velocity $V_D$. In Fig.~\ref{fig:conv} we plot aggregation time vs
$V_D$ for three different values of $n_0$. One observes that if
increasing the drift velocity from $V_D=0$, the aggregation time
decreases monotonically, with a steeper decline for larger $n_0$.
However, it soon ``saturates", so that increasing $V_D$ further
has very small effect on the aggregation time. This is because for
large values of $V_D/D_n$, the aggregation time is mainly
dominated by time required for robots to diffuse and detect
chemical gradient, and increasing $V_D$ clearly does not have any
effect on this time. Hence, depending on the desired number of
robots in the vicinity of the target, as well as the required
aggregation time, the best strategy for robots might be to have a
moderate drift velocity. Note that this type of analysis can be
used to assess the energy--efficiency of various behaviors since
power required to propel a robot through a fluid with velocity
$V_D$ scales with $V_D$.

.

\section{Discussion}

Microscopic robots face physical environments quite different from
our experience with larger robots. In particularly, many
interesting tasks for such robots will involve interaction with
spatially varying fields, such as chemical concentrations. In this
paper, we showed how to incorporate such fields into a previously
proposed approach relating individual robot behaviors to that of
the population as a whole. The analytic framework readily
incorporates Brownian motion and viscous fluid flow. The framework
also applies to macroscopic robots programmed to emulate behavior
of microscopic organisms --- for example,
chemotaxis~\cite{dhariwal04}.

As future work, it would be interesting to examine more realistic
scenarios, involving three-dimensional fluid flow in vessels with
quantitatively accurate physical parameters for the environment
and robot sensors and actuators. We could also compare alternate
approaches to various robot capabilities, e.g., acoustic vs.
chemical communication. For instance, with more extensive use of
communication, nearby robots could compare concentration
measurements to better estimate concentration gradients, thereby
introducing a spatial correlation in the state changes. Our
approach applies to these more complicated physical scenarios, but
becomes more computationally demanding to solve numerically.
Nevertheless, the computation cost involved to identify useful
design trade-offs through our approach will be less than that
involved with detailed simulations (especially for tasks involving
large populations of robots). Actual experiments with physical
devices will be even more challenging.

Limitations of this approach are its underlying assumptions,
namely that to determine relevant collective swarm behaviors the
occupancy numbers are a sufficient description of the individual
robots and fluctuations are small so the averages over many
independent instances of the swarm are close to the actual
observed behaviors in most of those instances. More broadly, this
approach connects average, aggregate behavior with local robot
controllers rather than providing specific details of individual
robots. For applications involving large populations of
microscopic robots, the law of large numbers will ensure behaviors
are usually close to average.

\small
\bibliographystyle{plain}
\bibliography{swarms}
\end{document}